\documentstyle[11pt,newpasp,twoside,epsf]{article}
\def\edcomment#1{\iffalse\marginpar{\raggedright\sl#1\/}\else\relax\fi}
\reversemarginpar

\begin{document}
\title{The Star Cluster Population of M51}

\author{Mark Gieles$^1$, Nate Bastian$^{1,2}$, Henny Lamers$^{1,3}$}
\affil{$^1$ Astronomical Institute, Utrecht University, Princetonplein 5,
3584 CC, Utrecht, The Netherlands\\
$^2$ European Southern Observatory, Karl-Schwarzschild-Strasse 2
D-85748 Garching b. M\"unchen, Germany\\
$^3$ SRON Laboratory for Space Research, Utrecht, The Netherlands}

\begin{abstract}
We present the age and mass distribution of star clusters in M51. The
 structural parameters are found by fitting cluster evolution models
 to the spectral energy distribution consisting of 8 {\em HST-WFPC2}
 pass bands. There is evidence for a burst of cluster formation at the
 moment of the second encounter with the companion NGC5195 (50-100 Myr
 ago) and a hint for an earlier burst (400-500 Myr ago). The cluster
 IMF has a power law slope of -2.1. The disruption time of clusters is
 extremely short ($< 100$ Myr for a 10$^4 M_{\odot}$ cluster).

\end{abstract}

\section{Burst(s?) in the cluster formation rate}
$N$-body models (Salo \& Laurikainen 2000) suggest 2 interactions
        between M51 and the companion NGC5195. The age distribution
        shows clear evidence for a burst $\sim$50-100 Myr ago and
        there is a hint for another burst at the epoch of the
        early encounter (see Fig.~1). The age distribution is complete
        for the last Gyr only for clusters with $M_{\rm cl} > 10^{4.7}
        M_{\odot}$ for which two bursts are clear. This makes it
        difficult to confirm the burst for all cluster masses.

\begin{figure}[!h]
\plotone{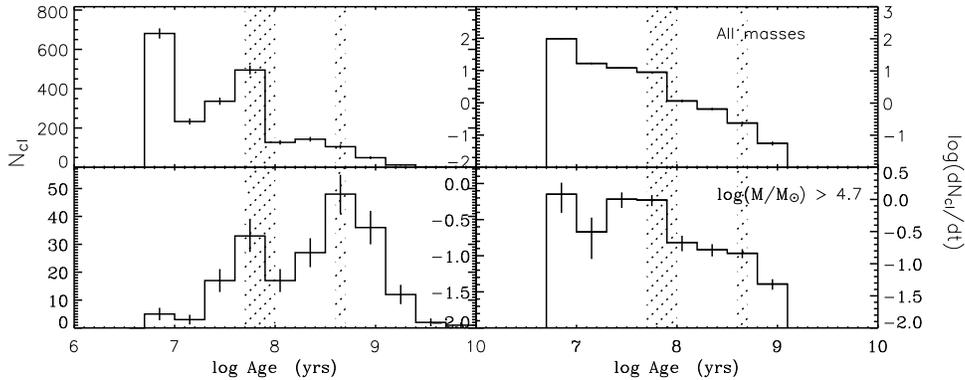}
\caption{Age histogram in absolute numbers (left) and in number of
clusters per Myr (right). The shaded regions are the predicted
moments of encounter with NGC5195.}
\end{figure}

\section{Disruption of clusters}


Analytical models (Lamers 2004) predicting the age and mass
distribution are fit to the data. The models assume: 1) a constant
cluster formation rate with a burst between 50-100 Myr; 2) a CIMF with
a slope of -2.1; 3) evolutionary fading under the detection limit and
4) a relation between the cluster disruption time and the cluster mass of the form: $t_{\rm dis}
= t_4 (M_{\rm cl}/10^4\>M_{\odot})^\gamma$ (Boutloukos \& Lamers 2003)

The value $\gamma$ is found to be 0.6 from observations and $N$-body
simulations (see Gieles et al., these proceedings). In Fig.~2 the best
fit models for the age and mass histogram are shown for all clusters
in M51 younger than 1 Gyr. The best fit value for $t_4$ is
$4\times10^7$ yr. To see whether different cluster disruption times
for different regions in M51 can be found, we divided M51 in rings
with a diameter of 1 kpc. For each ring the disruption rime $t_{4}$
was determined. Values between $2\times10^7$ yr and $10^8$ yr where
found, with no significant trend depending on the distance to the
galactic center.

\begin{figure}[!h]
\plotone{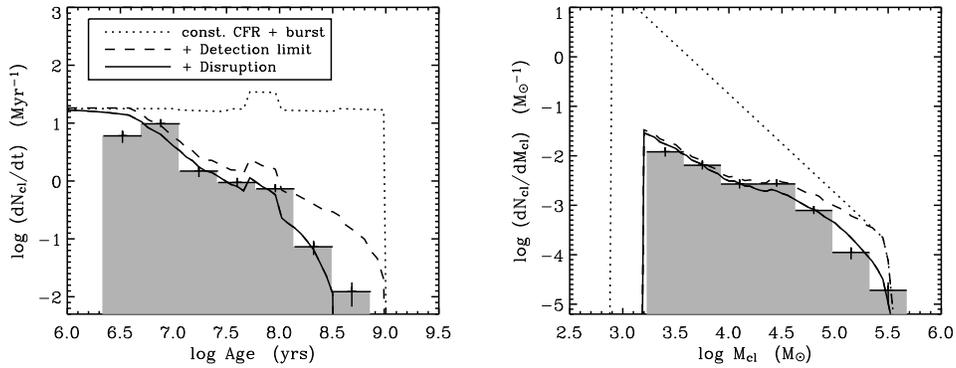}
\caption{Best fit models for age and mass distribution. A $t_4$ of
$4\times10^7$ Myr was needed to fit the observations. Left: number of
clusters per Myr. Right: number of cluster per $M_{\odot}$.}
\end{figure}


\begin{references}
\reference Boutloukos, S.~G. and Lamers, H.~J.~G.~L.~M., 2003, MNRAS 338, 717 
\reference Lamers,  H.~J.~G.~L.~M., 2004, MNRAS, in prep
\reference Salo, H., \& Laurikainen, E. 2000, MNRAS, 319, 377
\end{references}
\end{document}